# Interplay of Charge Density Wave States and Strain at the Surface of CeTe$_2$

Bishnu Sharma[1], Manoj Singh[1], Burhan Ahmed[1], Boning Yu[1], Philip Walmsley[2,3], Ian R. Fisher[2,3], Michael C. Boyer[1*]

[1]Department of Physics, Clark University, Worcester, Massachusetts 01610, USA

[2]Geballe Laboratory for Advanced Materials and Department of Applied Physics, Stanford University, Stanford, California 94305-4045, USA

[3]Stanford Institute for Materials and Energy Sciences, SLAC National Accelerator Laboratory, 2575 Sand Hill Road, Menlo Park, California 94025, USA

**Abstract:**
We use scanning tunneling microscopy (STM) to study charge density wave (CDW) states in the rare-earth di-telluride, CeTe$_2$. In contrast to previous experimental and first-principles studies of the rare-earth di-tellurides, our STM measurements surprisingly detect a unidirectional CDW with $q \sim 0.28\ a^*$, which is very close to what is found in experimental measurements of the related rare-earth tri-tellurides. Furthermore, in the vicinity of an extended sub-surface defect, we find spatially-separated as well as spatially-coexisting unidirectional CDWs at the surface of CeTe$_2$. We quantify the nanoscale strain and its variations induced by this defect, and establish a correlation between local lattice strain and the locally-established CDW states. Our measurements probe the fundamental properties of a weakly-bound two-dimensional Te-sheet, which experimental and theoretical work has previously established as the fundamental component driving much of the essential physics in both the rare-earth di- and tri-telluride compounds.

**Introduction:**

Charge density wave (CDW) states are found in numerous low-dimensional material systems where they coexist with other quantum orders such as superconductivity and magnetism. The interplay between elastic energy costs and electronic energy gains drive the details of a CDW state, characterized by charge localization, a periodic lattice distortion, and an energy gap in a material. Despite considerable progress in detailing these properties in a wide range of systems, fundamental questions persist. These questions range from determining the specifics of the driving mechanism for the CDW state within a given compound, to understanding the sensitivity of CDW states to parameters including elemental doping, external pressure, and strain.

---

\* To whom correspondence should be addressed: mboyer@clarku.edu

Here we present our studies on CeTe$_2$, a member of the rare-earth di-telluride compounds ($R$Te$_2$ where $R$ = rare-earth element). The $R$Te$_2$ compounds form in a Cu$_2$Sb-type tetragonal structure (space group P4/*nmm*) and are comprised by single Te square-planar sheets separated by rare-earth block layers (Figure 1a). The partially-filled in-plane 5$p_x$ and 5$p_y$ orbitals of the Te sheets allow for charge conduction within the *a-b* plane.[1] Separating each Te sheet is an insulating rare-earth block layer leading to a large out-of-plane resistivity and ultimately the quasi-two-dimensional nature of these materials. In CeTe$_2$, at room temperature, the CDW state is already well-established, and the resistivity along the *c*-axis is ~40 to 100 times that along the in-plane resistivity.[2,3]

First-principle calculations well-approximate the Fermi surface of the $R$Te$_2$ compounds mapped by angle resolved photoemission spectroscopy measurements,[1,2,4-7] indicating the primary contributions of the 5$p$ in-plane orbitals of the single Te square-planar sheets to the Fermi surface, and further illustrating the two-dimensional nature of these compounds. The parallel components of the Fermi surface indicate that the Fermi surface topology may be conducive to a Fermi-surface-nesting-driven CDW occurring in the $R$Te$_2$ materials. Indeed, initial transmission electron microscopy (TEM) measurements [4] detected a superlattice structure with $q_{CDW} \sim \frac{1}{2} a^*$ in the LaTe$_2$ compound in agreement with the initially-calculated Fermi surface nesting vector.[1,4] However, subsequent x-ray measurements [8] find variations in the measured $q_{CDW}$ including $q_{CDW} \sim \frac{1}{2} a^* + \frac{1}{2} b^*$ in LaTe$_2$, and TEM measurements [2] suggest a more-complicated series of CDW wavevectors in LaTe$_2$ and the closely-related CeTe$_2$. Thus the physics contained within these compounds is more-complex and less well-understood than initially believed. Further, Lindhard susceptibility calculations evince a range of possible wavevectors which could nest the $R$Te$_2$ Fermi surface.[2] Hence, the specific CDW nesting wavevector established within an $R$Te$_2$ material may be affected by a number of factors, including the choice of rare-earth ion, lattice strain, and elemental vacancies (e.g. Te vacancies) which can both effect band filling as well as introduce local lattice strain. Furthermore, while Fermi surface nesting has been identified as the primary candidate for the origin of the CDWs found in the $R$Te$_2$ compounds, we note that in the closely-related, and more intensely-studied $R$Te$_3$ compounds, the CDW-driving mechanism, initially suspected to be due to Fermi surface nesting, is now of debate. Both Fermi surface nesting and electron-phonon coupling mechanisms, or a combination of the two, have been identified as possible candidates.[8-14]

Given that considerably less attention has been given to the $R$Te$_2$ compounds, it is possible that the CDW-driving mechanism in these materials is not as clearly established as initially believed. Indeed, a strong electron-phonon coupling mechanism for the CDW in LaTe$_2$ compound has recently been suggested.[15]

In this paper, we present, what we believe to be, the first scanning tunneling microscopy (STM) measurements on near-stoichiometric CeTe$_2$ compounds. One goal of our studies is to image, in real space, the CDW state hosted by CeTe$_2$, and to compare our extracted $q_{CDW}$ to those previously found in the $R$Te$_2$ compounds as well as to that found in the related $R$Te$_3$ compounds. Given the expected sensitivity of $R$Te$_2$ compounds to external factors such as strain [2,16], and given the demonstrated sensitivity of CDW states in $R$Te$_3$ to local [17,18] and global strain [14,19,20], a second goal is to provide a nanoscale view of the interplay of local lattice strain and the locally-established CDW state(s) in CeTe$_2$.

**Experimental Methods:**

Single crystal CeTe$_2$ samples were grown using a self-flux technique described in detail elsewhere.[2] The growth technique used effectively minimizes Te vacancies in the Te-plane leading to near-stoichiometric samples. Unlike the $R$Te$_3$ compounds which easily cleave between the van-der-Waals-bonded neighboring double Te planes (Figure 1b), the $R$Te$_2$ compounds cleave between the more-tightly-bound single Te plane and the neighboring block layer as seen in Figure 1c. An additional, but minor, consideration is that during the crystal growth process, a very thin surface layer of CeTe$_3$ grows on top of the CeTe$_2$ bulk crystal. To ensure removal of the CeTe$_3$ surface layer, we first cleave our CeTe$_2$ samples in ambient conditions using a razor blade or by mechanically striking a cleave post glued to the surface using silver epoxy. Subsequently, samples are inspected to ensure that the sample cleave occurred deep within the original sample before the sample is inserted into the vacuum chamber. Our scanning tunneling microscopy measurements were conducted at ~300 K, a temperature at which CeTe$_2$ is already deep within the CDW state [2], and in ultra-high vacuum (~$10^{-9}$ Torr) using an RHK PanScan STM using a chemically etched tungsten tip. After chemical etching, the tungsten tip was annealed in-situ then sharpened through electron bombardment.

**Results and Discussion:**

**Surface Topography**

Given that the crystal structure of CeTe$_2$ is comprised of repeating an alternating pattern of single Te layers and rare-earth block layers, it is expected that there is an equal possibility of exposing either as the surface layer upon cleaving. Unlike our previous work on the related TbTe$_3$ compound, we found it considerably more difficult to find large-scale, atomically-flat regions on the exposed surfaces of CeTe$_2$. Often, our topographic images show a surface layer with small-scale step edges with hints of short-range atomic structure (Figure 2a) rather than large-scale atomically-flat regions with extended atomic structure which could be linked to the bulk crystal structure. The proliferation of step edges are likely due to the stronger bonding between neighboring layers within the $R$Te$_2$ compounds making cleaving more difficult; whereas $R$Te$_3$ compounds can easily be cleaved using tape, effective cleaving of CeTe$_2$ necessitates using a razor blade or mechanically hitting a cleave bar which is firmly fixed to the sample.

However, through large-scale scanning, we are indeed able to locate extensive, atomically-flat regions for imaging and study. Figure 2b shows a typical topography acquired across an atomically-flat region. Evident in the image are atomic periodicities over which a striped CDW pattern is superimposed. Figure 2c shows the Fast Fourier transform (FFT) of such a typical topography, which evinces peaks which we can associate with the crystal lattice and established CDW state. We identify four peaks (circled in blue) associated with a square lattice and with a periodicity consistent with the CeTe$_2$ bulk lattice parameter of $a = 4.47$ Å reported by x-ray measurements [21]. Additionally, we identify four peaks (circled in orange) associated with a square lattice rotated 45° with respect to this first lattice, and with periodicity of $\frac{a}{\sqrt{2}}$. Following our previous work on the related TbTe$_3$ compound, where a near-identical FFT lattice peak layout is observed, we identify the peaks circled in orange as originating from the Te layer, and the peaks circled in blue as originating from the rare-earth block layer. In short, when scanning these regions in CeTe$_2$, as with the $R$Te$_3$ compounds, the tunneling current is comprised of contributions originating from both the Te and the rare-earth block layers. Previous STM work on the related CeTe$_3$ suggests that the signal from the rare-earth block layer is dominated by the Ce ion.[22]

Furthermore, we identify four peaks (circled in yellow) associated with the striped CDW pattern. Figure 2d shows a linecut taken through the FFT starting at the origin, through these four CDW-associated peaks and extending past the block layer peak. To extract the peak locations, we fit Gaussians to the peaks in Figure 2d, determining the four peak locations as 0.276 $a^*$, 0.434 $a^*$, 0.566 $a^*$, and 0.720 $a^*$ respectively, with estimated uncertainties extracted from the variance-covariance matrix of 0.006 $a^*$, 0.002 $a^*$, 0.002 $a^*$, 0.002 $a^*$. These CDW-associated wavevectors are very close to those determined by STM measurements on the related $R$Te$_3$ compounds. Consequently, we label the peaks in Figure 2d as "2", "3", "4", and "5", given that their wavevector values are close to the ~2/7, ~3/7, ~4/7, and ~5/7 $q_{atom}$ CDW-associated wavevectors reported for STM measurements of the RTe$_3$ compounds [17,22,23]. Further, and following our previous work on TbTe$_3$, we identify peak "2" as the CDW wavevector, $q_{cdw}$, and peak "4" as the first harmonic, $2q_{cdw}$. Peaks "3" and "5" occur at $q_{atom} - 2q_{cdw}$ and $q_{atom} - q_{cdw}$ respectively, and we identify these peaks as resulting from wavevector mixing, the nature of which is discussed in detail elsewhere [17,22,23]. As such, the origin of these four peaks is consistent with an incommensurate, unidirectional CDW established on the surface of CeTe$_2$.

    Our identification of a unidirectional CDW in CeTe$_2$ is initially surprising. Previous TEM experiments report a considerably more-complex CDW state in the bulk of CeTe$_2$ characterized by five wavevectors.[2] Even in the case of the related LaTe$_2$ where a unidirectional CDW was detected [4], the wavevector of $q_{CDW} \sim 0.50\ a^*$ noticeably differs from any of the four CDW-associated wavevectors we report. Rather, the origin of the four wavevectors can be understood in terms of the structural relation CeTe$_2$ has to the $R$Te$_3$ compounds and the fundamental physics contained within a single Te sheet. Cleaving an $R$Te$_3$ compound occurs between neighboring Te planes (Figure 1b); the exposed surface is a single Te plane which is directly above a rare-earth block layer. In contrast, cleaving an $R$Te$_2$ compound occurs between a single Te plane and a rare-earth block layer; this gives a 50% chance that the exposed surface is a single Te plane which is directly above a rare-earth block layer. In short, in the case that the $R$Te$_2$ compound cleaves such that the exposed surface is a Te plane, the STM tip probes a surface configuration which is identical to that of the $R$Te$_3$ compounds. Our previous work on TbTe$_3$ illustrates that the surface Te layer is only weakly bound to the bulk below and, as such, can host CDW states which differ from that observed in the bulk.[17] As a consequence, when the surface terminal layer is a Te plane, STM measurements probe the fundamental physics contained within a single Te plane, but the plane is subjected to outside influences such as local strain fields. Interestingly, while a weakly-bound surface Te layer atop a block layer may appear physically similar for both the $R$Te$_2$ and $R$Te$_3$ compounds, one might still expect slightly different band-fillings between the

two [2] which would lead to differing Fermi surfaces. Given that the CDW wavevectors we detect in CeTe$_2$ are very close to those detected by STM in the $R$Te$_3$ compounds, differences in band-filling do not appear to significantly affect the detected CDW wavevectors. This may indicate that the observed CDW wavevectors are principally selected via a strongly momentum-dependent electron-phonon coupling rather than Fermi surface nesting, connecting to other recent work [12,24].

Given that there is only a 50% chance of exposing a Te plane when cleaving an $R$Te$_2$ compound, it is equally likely that a block layer will be exposed. If a block layer is exposed and if it is well-coupled to the bulk below, then it is possible that probing the block layer with STM may give insight into the bulk CDW. However, the block layer is insulating since rare earth ions in the block layer donate electrons to fill the p-orbitals in the Te plane.[1] As a consequence, it is likely difficult to image this layer. Indeed we have acquired numerous images of the CeTe$_2$ surface where neither large-scale (100s of nanometers) or atomic-scale features are able to be resolved; this may be attributed to scanning a surface insulating block layer. When nanoscale features and atomic resolution are obtained, we detect a unidirectional CDW consistent with a surface Te layer. In none of our images, which include just under 5,000 topographic images taken on multiple CeTe$_2$ crystals (as well as multiple cleaves), do we observe CDW wavevectors reflecting those previously reported in bulk studies of $R$Te$_2$ compounds. Finally, while there is a propensity for Te vacancies to be introduced into the Te plane during the crystal growth process of $R$Te$_2$ compounds [2], we do not typically observe atomic vacancies in our topographies. This appears to confirm the near-stoichiometric quality of our crystals.

## **Surface Strain and CDW States**

Properties of CDW states can be manipulated by straining a material. The application of external pressure or chemical pressure can strain a material's crystal lattice leading to changes in bulk-CDW transition temperatures as well as the emergence of other orders, including coexisting CDW states.[19,20] STM measurements are particularly useful in obtaining a local view of CDW states which, in turn, can be related to local lattice strain fields. For example, in NbSe$_2$, STM measurements have allowed for the identification of at least four different CDW orders which have been attributed to differing local lattice strain conditions.[25,26] In 1T-TiSe$_2$, STM measurements show that Cu intercalation leads to the formation of striped CDW order instead of the typical 2 x 2 CDW order observed without intercalation; strain was suggested as a possible mechanism.[27] Our recent work on TbTe$_3$ [17] demonstrates that the material surface can host at least two distinct CDW states. In TbTe$_3$, these states can exist separately or coexist with one another. Furthermore, we suggested that the specific local CDW order may be tied to the local strain field induced onto the crystal surface when the sample is cleaved.

The topographic image in Figure 3a was acquired over a 316 Å x 395 Å region of CeTe$_2$. Evident in the region, by eye, is a dominant unidirectional CDW state along the $a_2$ crystal axis in the lower left corner of the image, a dominant unidirectional CDW state along the $a_1$ axis in the upper right corner, and an apparent cross-over region in between. While the in-plane axes of bulk CeTe$_2$ crystal structure are equivalent, here we label the axes as $a_1$ and $a_2$ so as to provide clarity in our analysis of the CDW states contained within this topographic region. Also prominent in this region, and possibly driving the observed CDW evolution across the region, is an extended sub-surface defect of unknown origin leading to a ~3 nm high hill-like feature, where the top right part of the surface of the image is higher than that in the flat region in the lower left corner. In the vicinity of this defect, it is important to emphasize that the surface layer is continuous throughout the imaged region; there are no breaks or step edges, leading to the conclusion that the defect is sub-surface. As such, there is an evolution and coexistence of more than one CDW state within the imaged surface layer.

Fourier transforms of sub-regions denoted in red and blue in Figure 3b confirm the multiple CDW states contained within the larger topographic region. The FFT of the red sub-region (Figure 3c), a region which extends across the hill-like defect and slightly beyond, shows evidence for CDWs established along both $a_1$ and $a_2$ crystal axes. The FFT of the blue sub-region (Figure 3d), a region which is primarily in an atomically-flat region and extends only minimally onto the defect, shows a CDW predominately along the $a_2$ axis within the region. In particular, the FFT in Figure 3d strongly resembles that seen for a typical topography in Figure 2d. In both FFTs, peaks can be contributed to Te ions, Ce ions from the rare-earth layer, and to a unidirectional CDW; this is consistent with probing a Te layer on top of a block layer in the regions imaged in Figure 3. Given the obvious spatial variation in the CDW states hosted in this topographic region, we are led to the following questions: 1) How does the lattice evolve across the region? 2) How do the CDW states evolve across the region? 3) Is there a correlation between the lattice evolution and the locally-established CDWs?

To address these questions, we sub-divide the large topographic image into twenty 79 Å by 79 Å square regions (white squares in Figure 3b) and analyze each. Taking Fourier transforms of each allows for the determination of the associated local average lattice parameters and CDW wavevectors. Whereas in the our previous work we used the surface Te ion locations to determine the lattice parameters in the Te surface layer of TbTe$_3$, here we use the slightly-more prominent sub-surface-originating Ce (block-layer) peaks in the FFTs so as to extract the local lattice parameters. The lattice parameters were extracted by fitting Gaussians to the two Ce ion-associated peaks in each region's FFT to determine their associated wavevectors. As shown previously, due to the STM tip condition and/or variations in the coupling of the top Te layer to the block layer below, the intensity of the block-layer (Ce) peaks in the FFT can be greater than those of the Te layer, even if the Te layer is closest to the STM tip.[17] Furthermore, the ~1% smaller average lattice parameter in the Te layer (compared to the bulk lattice parameter found in TbTe$_3$ [17]) is insignificant as compared to the extreme lattice parameter variations found in this region of CeTe$_2$. Consequently, the use of the sub-surface Ce ions to determine the local lattice parameter is a good reflection of the lattice parameters and their variations, for the directly-imaged surface Te layer.

Plots in Figure 4a and 4b show the average $a_1$ and $a_2$ lattice parameters for each of the 20 square sub-regions and provide a visualization of the spatial variation for each and a connection to local strain. Both plots show localized regions with average lattice parameters which differ from that in the bulk, but the $a_1$ lattice parameter has a much stronger spatial variation, ranging from a minimum of 3.90 Å (13% compressive strain) to a maximum of 5.41 Å (21% tensile strain), than that of $a_2$, ranging from a minimum of 4.45 Å (~bulk value) to a maximum of 4.88 Å (9% tensile strain). Further, there is a clear spatial evolution to the lattice parameters where, on average, both $a_1$ and $a_2$ lattice parameters are smaller in the flat lower left of the region of Figure 3a compared to the defect region in the upper right. The sub-surface defect introduces significant lattice strain leading to an expansion of both lattice parameters in the defect-affected region. In short, the topographic region in Figure 3a provides an opportunity in which to study the nanoscale interplay of compressive and tensile strain and the locally-established CDW states in CeTe$_2$.

Figure 4c shows the ratio of the $a_1$ to the $a_2$ lattice parameter for each of the 20 sub-regions. Superimposed on top of this plot are two black dotted lines separating three distinct

CDW regions. In the lower left region, only CDW peaks along the $a_2^*$ axis are observed in FFTs of the sub-regions (no CDW peaks along the $a_1^*$ axis are observed above the noise level); this region has a unidirectional CDW established along the $a_2$ axis. In the upper right region, only CDW peaks along the $a_1^*$ direction are detected; this region has a unidirectional CDW established along the $a_1$ axis. In the middle region CDW peaks are observed along both the $a_1^*$ and $a_2^*$ directions; this region is a cross-over region where both unidirectional CDWs coexist.

The local evolution of the CDW states across these three regions follows the spatial evolution of the lattice strain. In the lower left region, there is compressive strain along the $a_1$ axis whereas the $a_2$ lattice parameter is close to that of the bulk. Throughout this region the $a_1$ lattice parameter is smaller than that of the $a_2$ lattice parameter, and a CDW state purely along the $a_2$ axis is established. In the upper right region, there is tensile strain along both the $a_1$ and $a_2$ axes; however, the tensile strain along the $a_1$ axis is larger, leading to $a_1$ lattice parameters which are larger than $a_2$ throughout the region. Within this region, only a unidirectional CDW along the $a_1$ axis is established. Finally, the middle region is a crossover region where both CDWs coexist, and neither the $a_1$ lattice parameter nor the $a_2$ lattice parameter is consistently larger. There appears a clear correlation between strain along the $a_1$ and $a_2$ axes and the established CDW states.

Next, we examine the local wavevectors, $q_{CDW1}$ and $q_{CDW2}$, associated with the CDW states established along the $a_1$ and $a_2$ axes respectively for each of the 20 sub-regions (Figures 5a and 5b). The regions in gray indicate the absence of the CDW along that direction; any CDW-associated peak in the FFT is within the noise level. In Figure 5, $q_{CDW1}$ and $q_{CDW2}$ are expressed in terms of the locally-associated $a_1^*$ and $a_2^*$ for an individual sub-region. In the region in Figure 2b, away from any obvious defects, $q_{CDW} = 0.28\ a^*$. Note that if the local CDW wavevectors within a given sub-region were pinned to the average local lattice parameter, the plots in Figures 5a and 5b would show a single white color throughout, indicating that the CDW wavevectors directly mimic changes to the average lattice parameter from one region to the next. Instead we find variations for each. Throughout the region, $q_{CDW2}$, shows relatively small variations (~0.27 – 0.29 $a_2^*$) as compared to $q_{CDW1}$. $q_{CDW1}$ ranges from 0.28 – 0.29 $a_1^*$ in the upper-right region in which there is only a CDW along the $a_1$ axis. Within the cross-over region, $q_{CDW1}$ varies significantly with a minimum of 0.26 $a_1^*$ in the top left-most corner and a maximum of 0.40 $a_1^*$ near-center. In general, in regions where there is only a single unidirectional state present, $q_{CDW1,2}$, is close to the 0.28 $a^*$ found in regions away from defects. In the crossover region, there are stronger variations in $q_{CDW1,2}$, though the CDW states remain collinear with the $a_1$ and $a_2$ axes across the entire topographic region.

As is evident in the top left of Figure 3a, the two unidirectional CDW states established in the region do not appear perpendicular to one another throughout, and appear, at some locations, to be rotated relative to crystal axes drawn as a guide on the figure. However, these drawn crystal axes are slightly misleading as the lattice unit cell changes throughout the region; the local $a_1$ and $a_2$ axes not only expand/contract, but also slightly rotate from one region to the next, as seen in Figures 6a and 6b. Figure 7a shows the *relative* angle between the $a_1$ and $a_2$ axes for each sub-region. For each sub-region the relative angle between the $a_1$ and $a_2$ axes is smaller than $90^0$, ranging from a minimum of $73^0$ to a maximum of $85^0$. Together, this indicates that in addition to compressive and tensile strain throughout the region, there is also lattice shear strain. This lattice shear strain affects the local CDW state by causing a rotation of the CDW wavevectors; the CDW remains collinear with the average local $a_1$ and $a_2$ axes for each of the sub-regions.

Figure 7b provides a visual illustration of the average unit cells for the two sub-regions of Figure 7a enclosed by dotted squares, and helps to highlight some of the longitudinal and rotational changes to the $a_1$ and $a_2$ axes as well as the unit cell which occur in the region. The first sub-region (dotted-dark blue square of Figure 7a) is in a region where there is purely a unidirectional CDW along the local $a_2$ axis; here the lattice parameters are such that $a_2 > a_1$. The second sub-region (dotted-red square in Figure 7a) is in a region where there is only a unidirectional CDW along the $a_1$ axis; here $a_1 > a_2$. The unit cells are drawn so as to preserve their relative scale and are oriented with respect to the x- and y-image scan directions. The rotation of the local $a_1$ and $a_2$ axes (e.g. unit cell) is clear, as is the ~$3^0$ relative angular change between the axes in the two regions.

**Conclusions:**

We have acquired the first STM measurements on a member of the $R$Te$_2$ compounds. In contrast to bulk measurements, we observe a unidirectional CDW with a wavevector very similar to that found in the $R$Te$_3$ compounds. This correspondence can be understood in terms of STM measurements directly probing a Te surface layer which is weakly bound to the rare-earth block layer below for each compound. Much of the essential physics within the R-Te compounds can be captured using a model of a single square Te plane.[4,10,11] As a consequence, our STM measurements probe the essential physics contained within this plane which is subject to weak coupling to the block layer and to lattice strain induced by defects. By studying changes in the lattice parameters and CDW states in the vicinity of an extended defect, we find a correlation between nanoscale lattice strain and locally-established CDW states.

**Acknowledgements:**
Scanning tunneling microscopy studies at Clark University were supported by the National Science Foundation under Grant No. DMR-1904918. Crystal growth and characterization at Stanford University (PW and IRF) was supported by the Department of Energy, Office of Basic Energy Sciences under contract DE-AC02-76SF00515.

**References:**
[1]    A. Kikuchi, Journal of the Physical Society of Japan **67**, 1308 (1998).
[2]    K. Y. Shin, V. Brouet, N. Ru, Z. X. Shen, and I. R. Fisher, Physical Review B **72**, 085132 (2005).
[3]    B. H. Min, E. D. Moon, H. J. Im, S. O. Hong, Y. S. Kwon, D. L. Kim, and H. C. Ri, Physica B: Condensed Matter **312-313**, 205 (2002).
[4]    E. DiMasi, B. Foran, M. C. Aronson, and S. Lee, Physical Review B **54**, 13587 (1996).
[5]    J. H. Shim, J. S. Kang, and B. I. Min, Physical Review Letters **93**, 156406 (2004).
[6]    E. Lee, D. H. Kim, J. D. Denlinger, J. Kim, K. Kim, B. I. Min, B. H. Min, Y. S. Kwon, and J. S. Kang, Physical Review B **91**, 125137 (2015).
[7]    E. Lee, H. W. Kim, S. Seong, J. D. Denlinger, Y. S. Kwon, and J. S. Kang, Journal of the Korean Physical Society **70**, 389 (2017).
[8]    R. G. Moore, V. Brouet, R. He, D. H. Lu, N. Ru, J. H. Chu, I. R. Fisher, and Z. X. Shen, Physical Review B **81**, 073102 (2010).
[9]    J. Laverock, S. B. Dugdale, Z. Major, M. A. Alam, N. Ru, I. R. Fisher, G. Santi, and E. Bruno, Physical Review B **71**, 085114 (2005).
[10]   E. DiMasi, M. C. Aronson, J. F. Mansfield, B. Foran, and S. Lee, Physical Review B **52**, 14516 (1995).
[11]   H. Yao, J. A. Robertson, E.-A. Kim, and S. A. Kivelson, Physical Review B **74**, 245126 (2006).
[12]   M. Maschek, S. Rosenkranz, R. Heid, A. H. Said, P. Giraldo-Gallo, I. R. Fisher, and F. Weber, Physical Review B **91**, 235146 (2015).
[13]   H.-M. Eiter, M. Lavagnini, R. Hackl, E. A. Nowadnick, A. F. Kemper, T. P. Devereaux, J.-H. Chu, J. G. Analytis, I. R. Fisher, and L. Degiorgi, Proceedings of the National Academy of Sciences **110**, 64 (2013).
[14]   R. G. Moore *et al.*, Physical Review B **93**, 024304 (2016).
[15]   H. X. Yang, Y. Cai, C. Ma, J. Li, Y. J. Long, G. F. Chen, H. F. Tian, L. L. Wei, and J. Q. Li, EPL (Europhysics Letters) **114**, 67002 (2016).
[16]   M. Lavagnini, A. Sacchetti, L. Degiorgi, E. Arcangeletti, L. Baldassarre, P. Postorino, S. Lupi, A. Perucchi, K. Y. Shin, and I. R. Fisher, Physical Review B **77**, 165132 (2008).
[17]   L. Fu, A. M. Kraft, B. Sharma, M. Singh, P. Walmsley, I. R. Fisher, and M. C. Boyer, Physical Review B **94**, 205101 (2016).
[18]   A. Fang, J. A. W. Straquadine, I. R. Fisher, S. A. Kivelson, and A. Kapitulnik, arXiv:1901.03471 (2019).
[19]   N. Ru, C. L. Condron, G. Y. Margulis, K. Y. Shin, J.Laverock, S. B. Dugdale, M. F. Toney, and I. R. Fisher, Physical Review B **77**, 035114 (2008).
[20]   D. A. Zocco, J. J. Hamlin, K. Grube, J. H. Chu, H. H. Kuo, I. R. Fisher, and M. B. Maple, Physical Review B **91**, 205114 (2015).
[21]   M. H. Jung, K. Umeo, T. Fujita, and T. Takabatake, Physical Review B **62**, 11609 (2000).
[22]   A. Tomic, Z. Rak, J. P. Veazey, C. D. Malliakas, S. D. Mahanti, M. G. Kanatzidis, and S. H. Tessmer, Physical Review B **79**, 085422 (2009).
[23]   A. Fang, N. Ru, I. R. Fisher, and A. Kapitulnik, Physical Review Letters **99**, 046401 (2007).
[24]   M. D. Johannes and I. I. Mazin, Physical Review B **77**, 165135 (2008).
[25]   A. Soumyanarayanan, M. M. Yee, Y. He, J. van Wezel, D. J. Rahn, K. Rossnagel, E. W. Hudson, M. R. Norman, and J. E. Hoffman, Proceedings of the National Academy of Sciences **110**, 1623 (2013).

[26]    S. Gao, F. Flicker, R. Sankar, H. Zhao, Z. Ren, B. Rachmilowitz, S. Balachandar, F. Chou, K. S. Burch, Z. Wang, J. van Wezel, and I. Zeljkovic, Proceedings of the National Academy of Sciences (2018).
[27]    A. M. Novello, M. Spera, A. Scarfato, A. Ubaldini, E. Giannini, D. R. Bowler, and C. Renner, Physical Review Letters **118**, 017002 (2017).
[28]    K. Momma and F. Izumi, Journal of Applied Crystallography **44**, 1272 (2011).

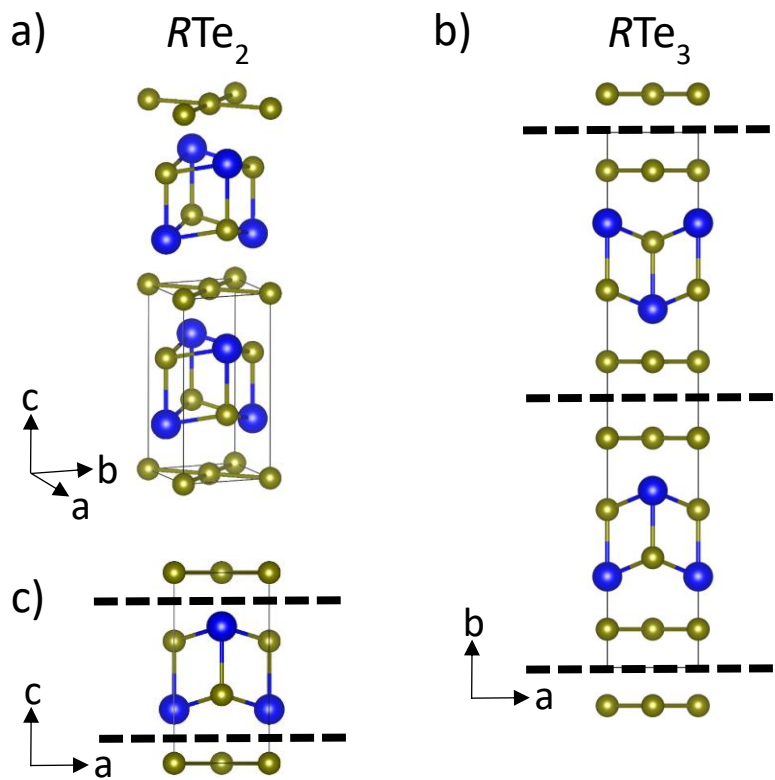

**Figure 1**: Ions in gold color are Te ions. Ions in blue are rare-earth ions. a) Crystal structure for $R$Te$_2$ compounds. b) Crystal structure for $R$Te$_3$ compounds. Crystal cleave planes (dotted black lines) are between neighboring square-planar Te sheets. Cleaving the crystal exposes a square-planar Te sheet surface layer. c) Two possible cleave planes (dotted black lines) for $R$Te$_2$. The top cleave plane would results in a surface rare-earth block layer. The bottom cleave plane would result in a surface Te sheet. The crystal structures were constructed using Vesta software [28].

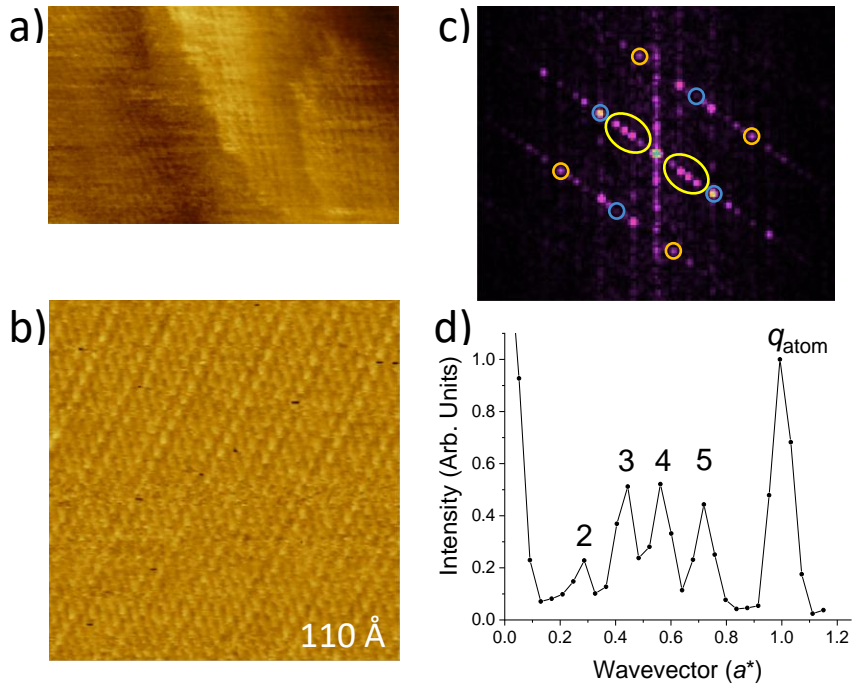

**Figure 2**: a) 100 Å x 60 Å topographic image showing step edges as well as hints of atomic structure. The image was acquired with $V_{sample}$ = +500 mV and I = 200 pA. b) Typical topography in a flat region of the sample surface. The unidirectional CDW is seen in stripes superimposed on an atomic structure. The image was acquired with $V_{sample}$ = +200 mV and I = 400 pA. c) FFT of a typical topography. Orange circles enclose peaks originating from the surface square-planar Te sheet. Blue circles enclose peaks originating from the sub-surface rare-earth block layer (Ce ions). Yellow ovals enclose 4 peaks associated with the unidirectional CDW and those originating from mixing between the CDW wavevector and the block layer atomic wavevectors. d) Linecut through the FFT in c) beginning at the origin, in the direction of the CDW, and extending just past the atomic signal originating from the block layer ($q_{atom}$).

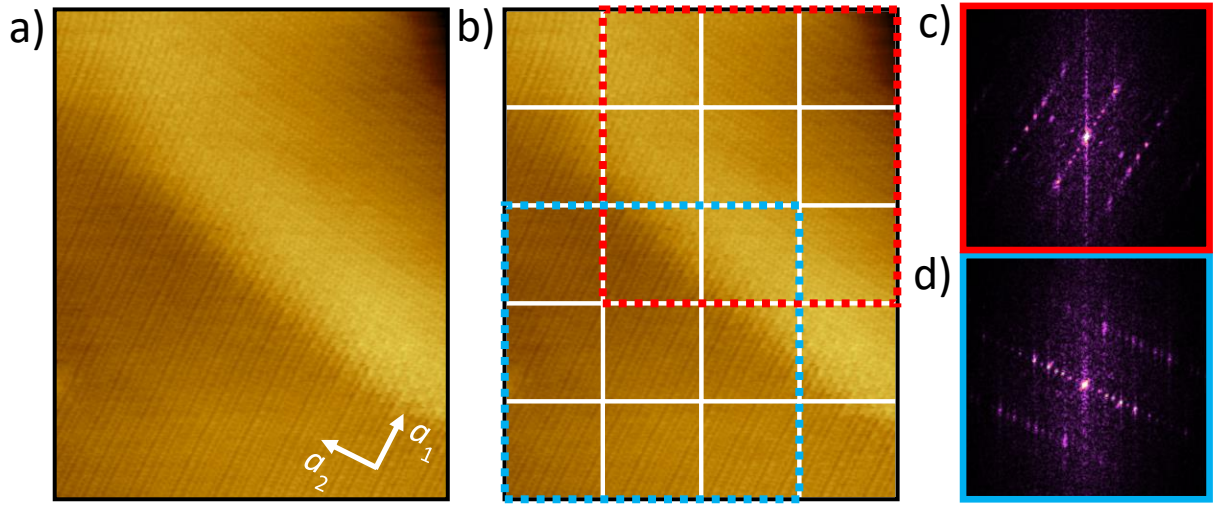

**Figure 3**: a) 316 Å x 395 Å topographic image showing two unidirectional CDWs, a cross-over region where the two CDWs coexist, and an extended sub-surface defect. The image was acquired with $V_{sample}$ = +150 mV and I = 500 pA. b) The topographic image in a) is split into twenty 79 Å x 79 Å sub-regions (white squares). Analysis of each of these sub-regions allows for a more-local understanding of the CDW states, local lattice parameters, and possible correlations between the two. c) FFT of region of b) enclosed by the red-dotted square. This region extends over much of the sub-surface defect and shows evidence for CDW states along both the $a_1$ and $a_2$ axes, in addition to atomic periodicities associated with both the Te layer and rare-earth block layer. d) FFT of region of b) enclosed by the blue-dotted square. This region is primarily in the atomically-flat region of the topographic image but its corner extends into the defect-affected area. Here only a unidirectional CDW state along the $a_2$ axis is obvious by eye. This FFT resembles the FFT in Figure 2c for that of a typical surface region on $CeTe_2$.

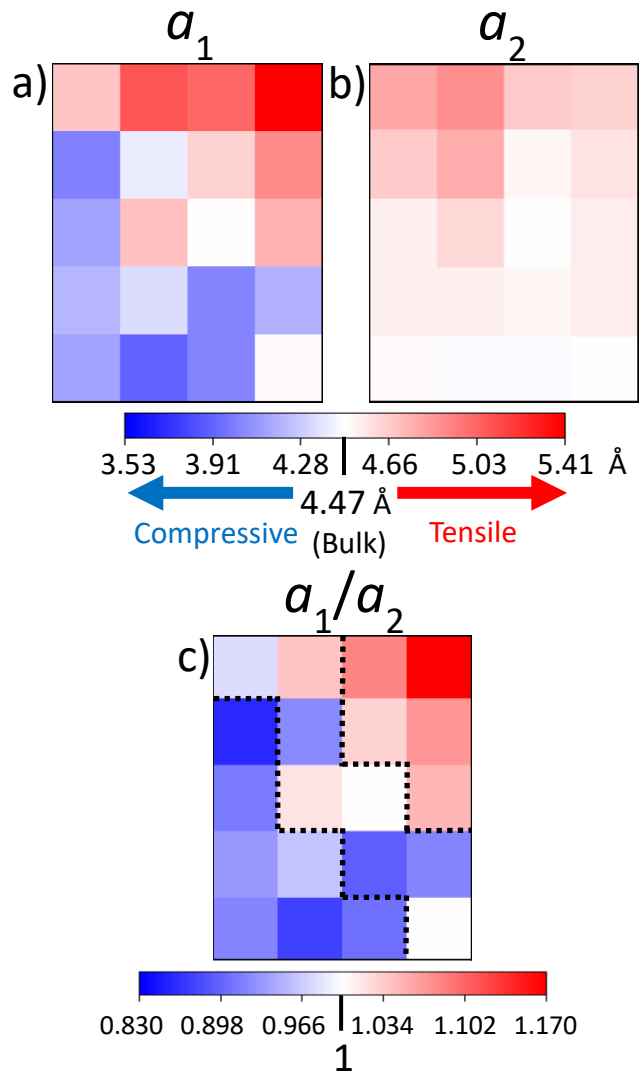

**Figure 4**: a) Plot of the $a_1$ lattice parameter for each of the 20 sub-regions of figure 3b. There is a strong spatial variation in the $a_1$ lattice parameter across the full region. In the atomically flat region, the lattice parameter is lower than that in the bulk, indicating compressive strain. In the defect-affected region, the $a_1$ lattice parameter becomes significantly larger than that of bulk indicating tensile strain. b) Plot of the $a_2$ lattice parameter for each of the 20 sub-regions of Figure 3b. The $a_2$ lattice parameter shows less variation across the topographic region than does the $a_1$ lattice parameter. However, in the defect-affected area, the $a_2$ lattice parameter becomes larger than that for the bulk indicating tensile strain. c) Plot of the ratio of the two lattice parameters ($a_1/a_2$) for each of the 20 sub-regions. Dotted lines are superimposed on the plot to indicate three regions. The lower left corner, where $a_1/a_2$ is always less than 1, hosts only a unidirectional CDW along the $a_2$ axis. The upper right corner, where $a_1/a_2$ is always greater than 1, is a region where there is only a unidirectional CDW along the $a_1$ axis. The region in between is a cross-over region where both unidirectional CDWs coexist, and neither the $a_1$ or $a_2$ local lattice parameter is consistently larger throughout this region.

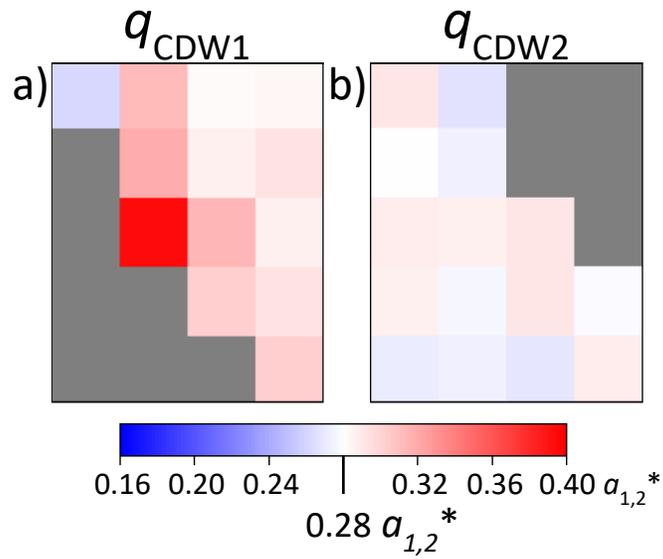

**Figure 5**: a) Plot of the wavevector for the unidirectional CDW along the $a_1$ axis ($q_{CDW1}$) for each of the 20 sub-regions. The grey region of the plot indicates a region where any signal associated with the $a_1$ axis CDW is absent (within the noise level). $q_{CDW1}$ and $q_{CDW2}$ (for Figure b) are expressed in terms of the locally-associated $a_1^*$ and $a_2^*$ for an individual sub-region. b) Plot of the wavevector for the unidirectional CDW along $a_2$ axis ($q_{CDW2}$) for each of the 20 sub-regions. The CDW along the $a_2$ axis is absent in the grey region.

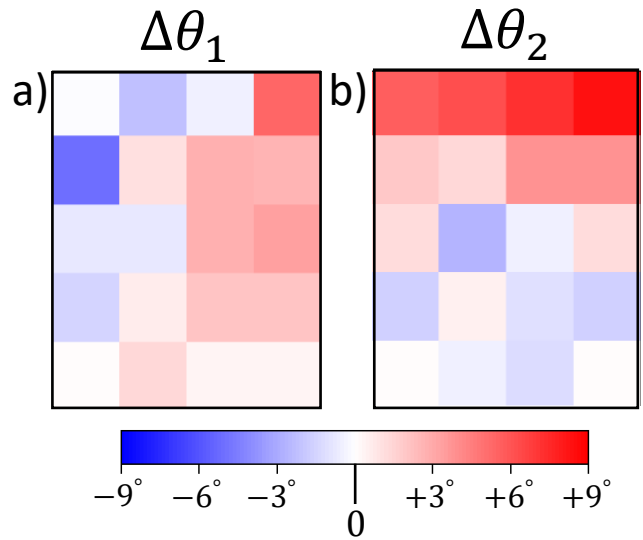

**Figure 6**: a) Angular rotation of the $a_1$ lattice parameter for each of the 20 sub-regions relative to its orientation in the sub-region in the lower left corner of the plot. "+" represents a counter clockwise rotation and "–" represents a clockwise rotation. b) Angular rotation of $a_2$ lattice parameter as compared to its orientation in the sub-region in the lower left corner of the plot.

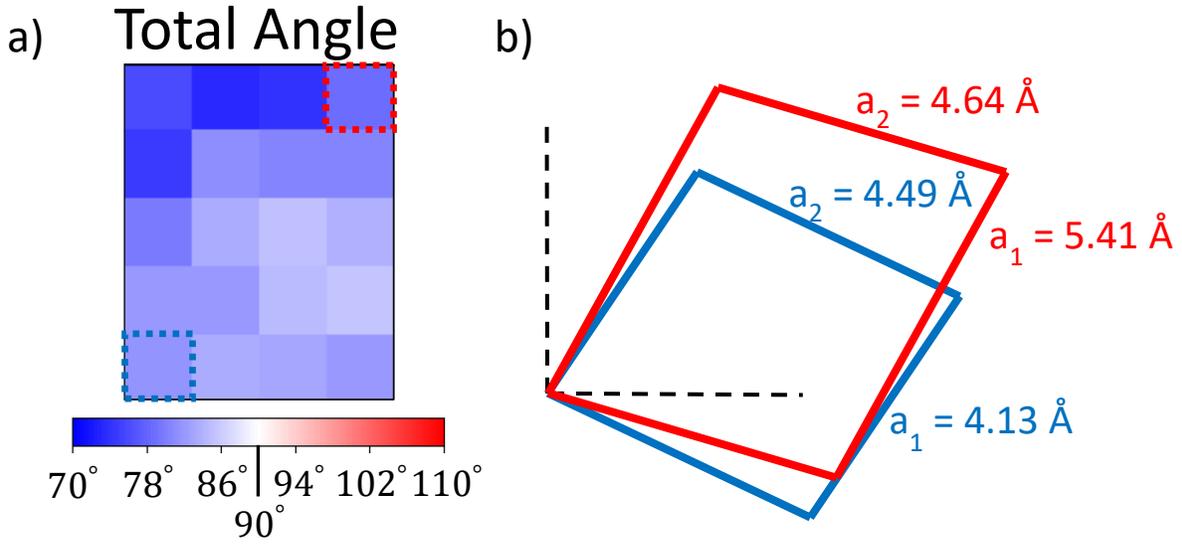

**Figure 7**: a) Plot of the total angle between the $a_1$ and $a_2$ lattice parameters for each of the 20 sub-regions. b) Shown in blue is the unit cell for the blue-dotted region in a). Here $a_1 < a_2$, and there is a CDW only along the $a_2$ axis. In red, the unit cell for the red-dotted region in a) is drawn preserving the scale and orientation relative to the blue unit cell. In this sub-region, $a_1 > a_2$ there is a CDW only along the $a_1$ axis.